\documentclass[a4paper,11pt]{article}
\pdfoutput=1 % if your are submitting a pdflatex (i.e. if you have
\usepackage{jinstpub} % for details on the use of the package, please
\usepackage{url}
\usepackage[version=3]{mhchem}

\newcommand{\Figure}{figure~}
\newcommand{\Table}{table~}
\newcommand{\JSNS}{JSNS$^2$\,}

\newcommand{\nuebar}{$\bar{\nu}_{e}$\,}

\newcommand{\Sterile}{$\bar{\nu}_{\mu} \to \bar{\nu}_{e}$~}

\newcommand{\IBD}{$\bar{\nu}_{e} + p \to e^{+} + n$~}

\newcommand{\msq}{$\mathrm{m}^2$\,}
\newcommand{\micro}{$\mu \mathrm{s}$\,}

\title{\boldmath Aging study of Gd concentration in LAB-based Gd loaded liquid scintillator exposed to passivated stainless steel}

\author[1]{Y.~Hino,\note{Corresponding author.}}
\author[2]{H.~Furuta,\note{now at High Energy Accelerator Research Organization (KEK), 1-1 Oho, Tsukuba, Ibaraki, Japan.}}
\author[]{and F.~Suekane}

\affiliation[]{Research Center for Neutrino Science, Tohoku University,\\6-3 Aramaki Aza-aoba, Aoba-ku, Sendai, Miyagi, Japan}

\emailAdd{hino@awa.tohoku.ac.jp}

\abstract{
    Stainless steel is a candidate of the inner surface of the storage container for LAB-based Gadolinium loaded liquid scintillator (Gd-LS) to be used in the \JSNS neutrino detector.
    Aging effect on Gd concentration of Gd-LS was investigated with the Gd-LS sample stored in the passivated stainless steel bottle in two independent methods.
    The direct comparison of the neutron capture time measurement showed that there is no significant degradation of the capture time after 602 days aging.
    Titration was performed to measure Gd concentration, and the result after 466 days aging is consistent with the result before aging within the uncertainty of the measurement. The upper limit of degradation of Gd concentration in 21 kL tank case is estimated as 0.5 \% for 10 years of storage.
    Both results lead to a conclusion that stainless steel is usable for Gd-LS storage for \JSNS experiment.
}

\keywords{Liquid Detectors, Neutrino Detectors}

\proceeding{Preprint typeset in JINST style}

\begin{document}
\maketitle
\flushbottom

\section{Introduction}
\quad \JSNS (J-PARC Sterile Neutrino Search at J-PARC Spallation Neutron Source) is an experiment to search for sterile neutrinos via the observation of \Sterile appearance oscillation \cite{Harada:2013}.
The detector contains 17 tons of Gadolinium (Gd) loaded liquid scintillator (Gd-LS) for a delayed coincidence detection of the inverse beta-decay (IBD) reaction \IBD. The positron yields scintillation light instantaneously regarded as a prompt signal. When the neutron is thermalized and captured by Gd, several gamma-rays with $\sim 8$ MeV total energy are emitted. The gamma-rays generate scintillation light observed as a delayed signal around 30 \micro after the prompt signal. The \nuebar signals are identified by the delayed coincidence of the prompt and delayed signals.

The Gd-LS is donated by Daya Bay~\cite{Ding:2008} group to \JSNS. It consists of LAB (linear alkyl benzene) as the base solvent, 3 g/L PPO (2,5-diphenyloxazole) as the fluor, and 15 mg/L bis-MSB (1,4-bis(2-methylstyryl) benzene) as the wavelength shifter. Gd is solved as a Gd carboxylate complex with 3,5,5-trimethylhexanoic acid (TMHA) ligands, and its concentration is 0.1 w\%. 
The \JSNS Gd-LS is planned to be stored in an ISO tank, a container based on ISO international standard, during annual MLF maintenance periods~\cite{Ajimura:2017}. The duration of storage in ISO tank is about two months a year.

It has been reported that there is no significant aging of essential properties of LAB-based LS, e.g., light yield, attenuation length, and Fe impurity, stored in a stainless steel container~\cite{Chen:2015}. The results are reasonable because stainless steel generally forms chemically inactive passivation film on the surface, which leads to high resistance to liquid organics.
Recently light yield stability of LAB-based Gd-LS has been reported in \cite{Gromov:2018}, and its result is consistent with the result shown in \cite{Chen:2015}. However, no experimental data about degradation of Gd concentration of LAB-based Gd-LS exposed to stainless steel has been reported.
Therefore, this paper focuses on an investigation of aging of Gd concentration in two different methods. One of them is a measurement of the neutron capture time as the practical property of Gd-LS for the IBD detection. The other is a titration to directly measure the Gd concentration via a chemical process.

The Gd-LS used in this study has the same composition as the Gd-LS for \JSNS experiment. 10 L of the Gd-LS sample has been stored in a passivated stainless steel (SUS304) bottle at room temperature. 
The area of the stainless steel surface contacting with the Gd-LS is roughly 0.25 \msq, and that of the ISO tank is $\sim 54$ \msq for 21 kL volume. Therefore, the storage in the stainless steel bottle is in 10 times more conservative situation than the ISO tank case in term of a ratio of the exposure area to Gd-LS volume.

\section{Neutron capture time measurement}

\begin{figure}[htbp]
    \centering 
    \includegraphics[width=.9\textwidth]{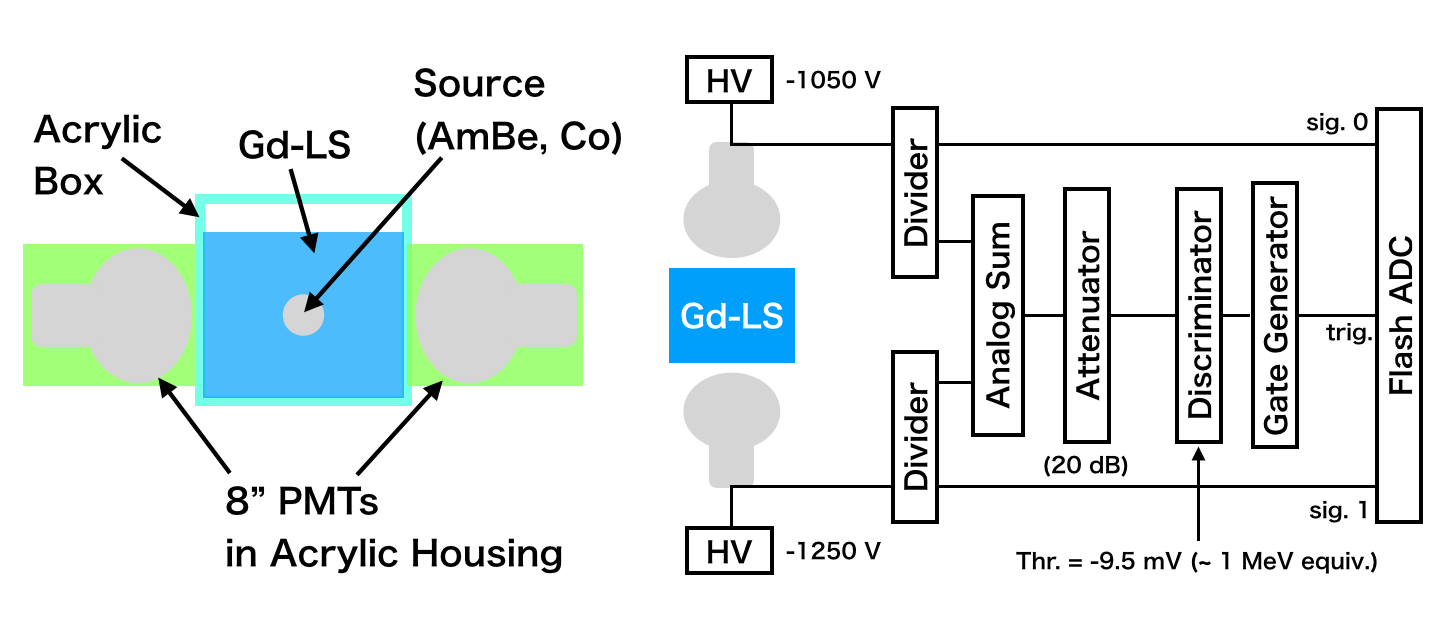}
    \caption{Left: a schematic view of the experimental setup for the capture time measurement. Right: Readout electronics. \label{fig:SetupCapT}}
\end{figure}

The experimental setup for the measurement of the neutron capture time is shown in \Figure \ref{fig:SetupCapT}. 10 L of the Gd-LS was contained in the acrylic box and two 8 inch photomultiplier tubes (PMTs) were placed at the opposite sides to detect scintillation light. 
At the center of the Gd-LS, an encapsuled Amerisium-Beryllium ($^{241}\mathrm{Am}{}^{9}\mathrm{Be}$), a gamma ray and neutron source, in a PTFE capsule was placed with a nylon string. 
IBD like signals are induced by the AmBe source, and are detected in the delayed coincidence method. The gamma ray and the fast neutron from AmBe generates scintillation light detected as a prompt signal. After the neutron is thermalized in the Gd-LS, the capture gamma rays are emitted from Gd, and makes a delayed signal. The capture time $\Delta t$ is defined as the time interval between a pair of the prompt and delayed signals.

The signals from PMTs were acquired in the system illustrated in the right of \Figure \ref{fig:SetupCapT} with a CAEN V1721 Flash Analog to Digital Converter (FADC) module which has 8-bit resolution and 500 MHz sampling rate. The trigger for data acquisition was generated by a discriminated signal from an analog summation of the two PMT signals. A pair of the prompt and the delayed signals was determined in off-line analysis sequence explained later.

\subsection{Analysis and Result}

To make a pair of the prompt and the delayed signals induced by the AmBe source, we defined prompt signal candidates and delayed signal candidates as follows. 
First we defined all triggered events as prompt signal candidates. A delayed signal candidate for a corresponding prompt signal candidate is defined as a signal detected within 1000 \micro after the prompt signal candidates. 
A signal observed between 1000 to 2000 \micro after the prompt candidate is used to estimate accidental background.
The pairs of the accidental signals were used for a statistical background subtraction.

\begin{figure}[htbp]
        \centering 
        \includegraphics[width=.45\textwidth]{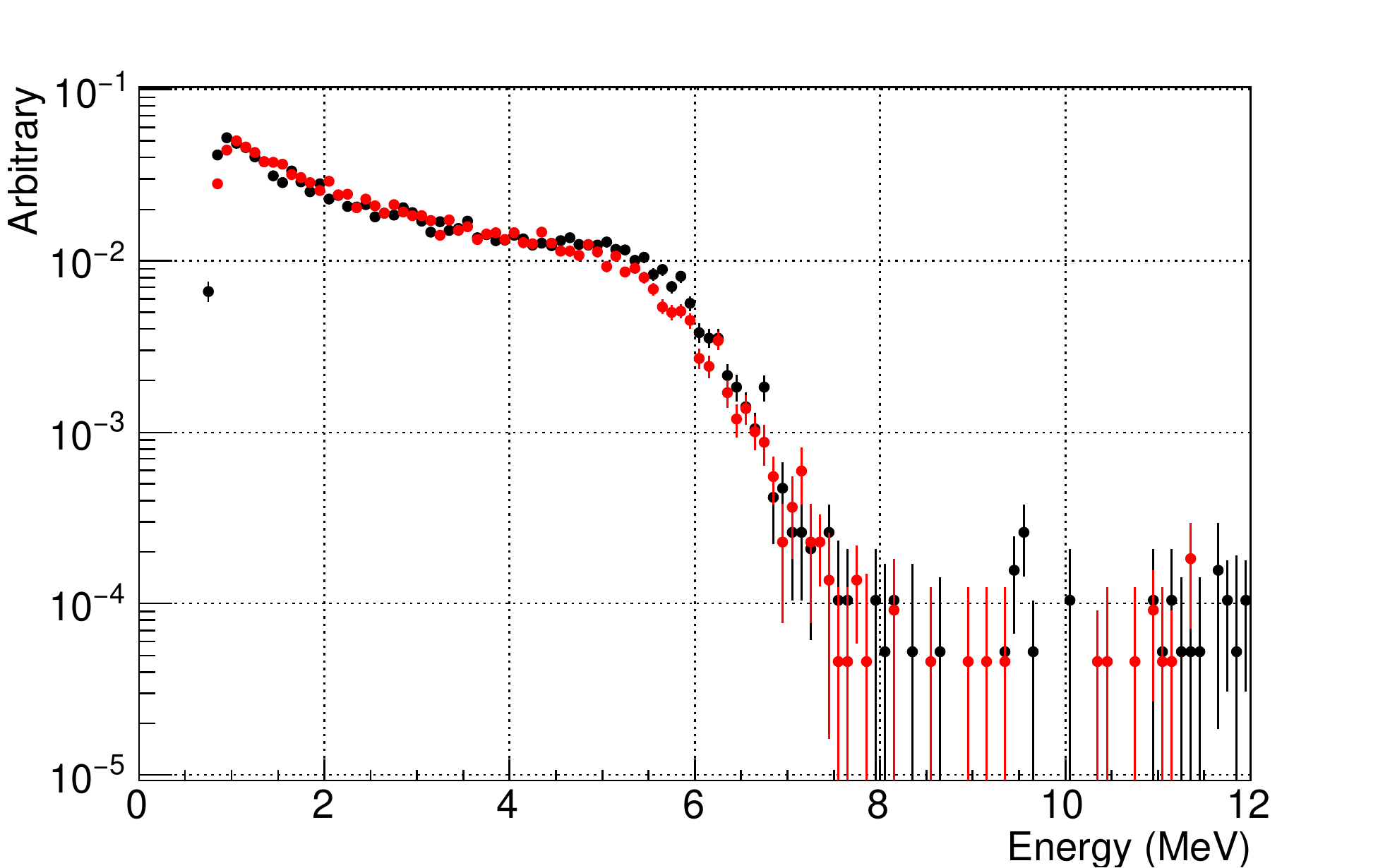}
        \quad
        \includegraphics[width=.45\textwidth]{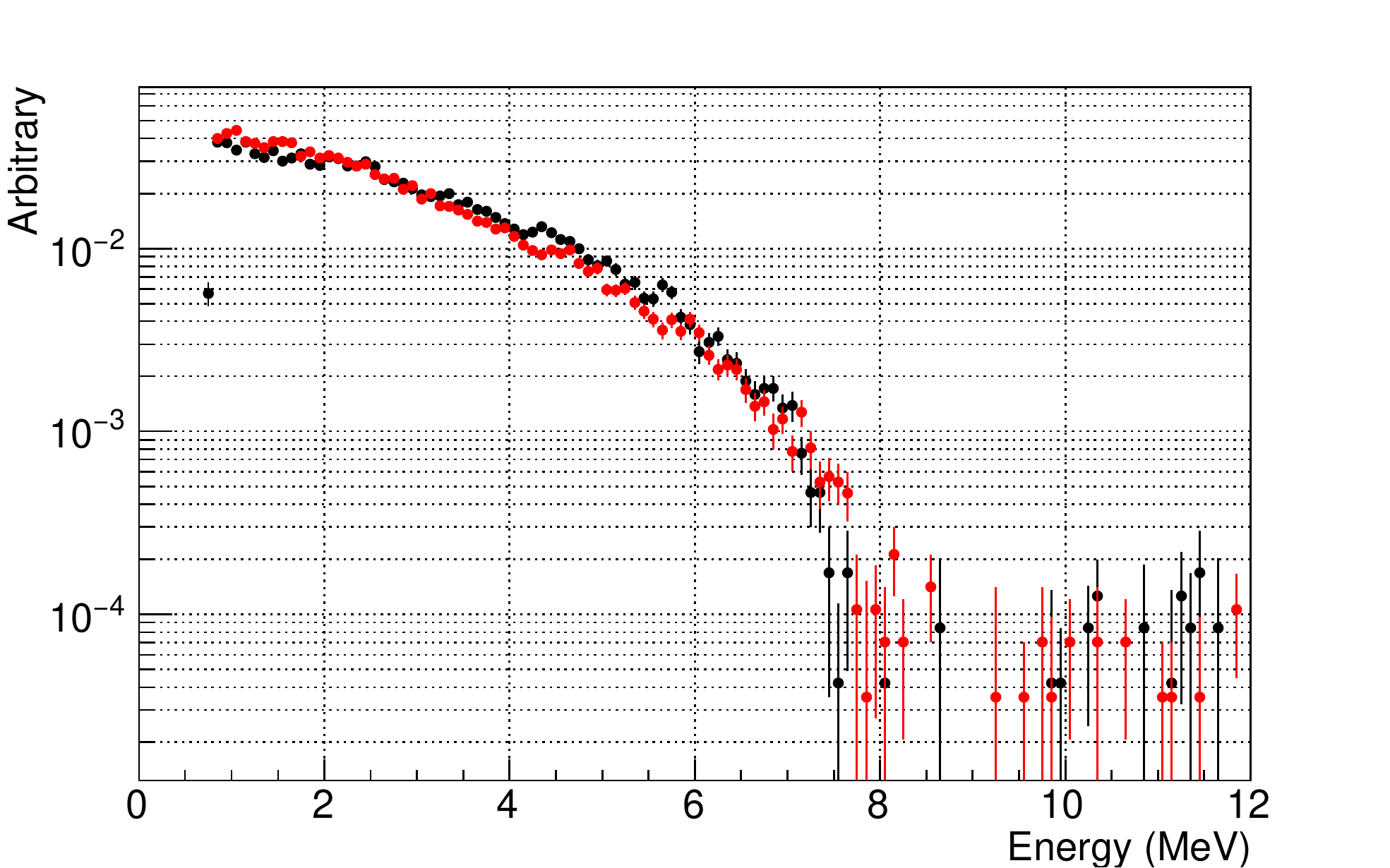}
    \caption{Plots of visible energy spectra of the prompt candidates (left) and the delayed candidates (right) after the accidental background subtraction. The black markers indicate the data before exposure to stainless steel. The red ones exhibit the data after 617 days aging. \label{fig:Evis}}
\end{figure}

\begin{figure}[htbp]
    \centering 
    \includegraphics[width=.5\textwidth]{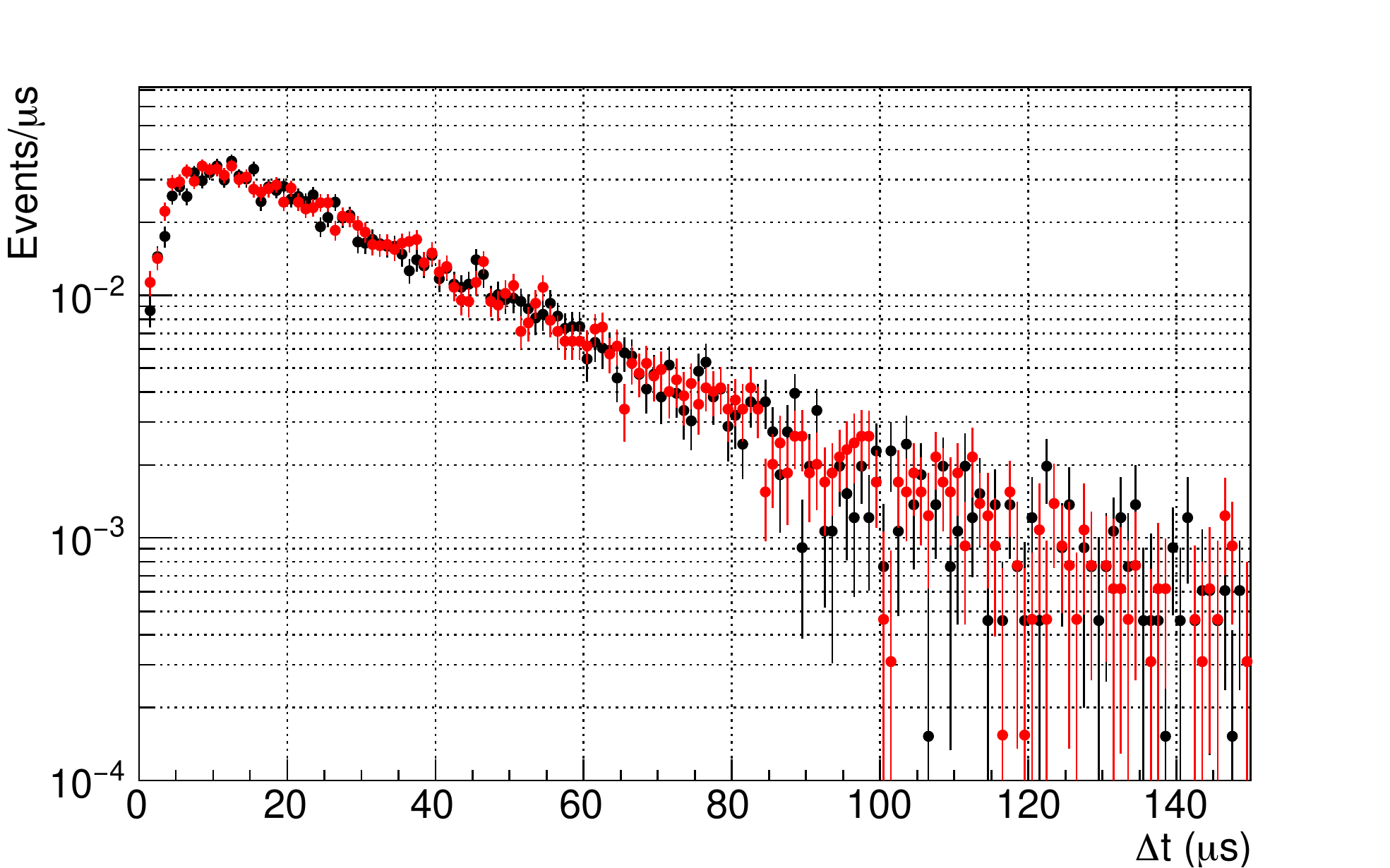}
    \caption{A plot shows a comparison of the $\Delta t$ distribution measured before aging (black) and after 617 days exposure (red). They well agree with each other. \label{fig:CapTcomp}}
\end{figure}

Figure \ref{fig:Evis} shows visible energy spectra of the prompt (left) and the delayed signal candidates (right) after the accidental background subtraction. The energy scale of the summation of the PMT charges was calibrated using ${}^{60}\mathrm{Co}$ source with replacing the AmBe source.
The spectra of the prompt candidates have a characteristic edge structure in 4 - 8 MeV region as a result of the superposition of energy deposits by the recoiled electron by compton scattering of the gamma-ray ($\sim 4$ MeV) and the recoiled proton by the fast neutron from AmBe. The delayed candidate spectra has no peak representing the total energy of the gamma-rays from Gd ($\sim 8$ MeV) due to a leak of the gamma-rays from the Gd-LS volume. However, the delayed signal candidates show spectra which have an end point at $\sim 8$ MeV.
Both the spectra measured before the stainless steel exposure (black) and 617 days after the exposure (red) show quite similar spectra in the entire range.
The Gd capture events for the capture time measurement were selected with the selection criteria listed in \Table \ref{tab:selection}. In order to collect events induced by AmBe source, the selection of the prompt candidates requires the energy range between 4 and 8 MeV. The lower end of the selection for the delayed signals was set to 4 MeV in order to reject the neutron capture events by hydrogen (2.2 MeV). 
Figure \ref{fig:CapTcomp} shows the capture time distributions. They show good agreement with each other.

\begin{table}[hbt]
  \caption{The selection criteria for AmBe events. \label{tab:selection}}
  \centering
    \begin{tabular}{|cccc|}
        \hline
        Event & Time Window/\micro & Prompt $E_{\mathrm{p}}$/MeV & Delayed $E_{\mathrm{d}}$/MeV \\ \hline
        nCapture & 0 - 1000   & 4 - 8 & 4 - 8  \\ %\hline
        Accidental & 1000 - 2000   & 4 - 8 & 4 - 8  \\ \hline
    \end{tabular}
\end{table}

\begin{figure}[htbp]
    \centering 
    \includegraphics[width=.45\textwidth]{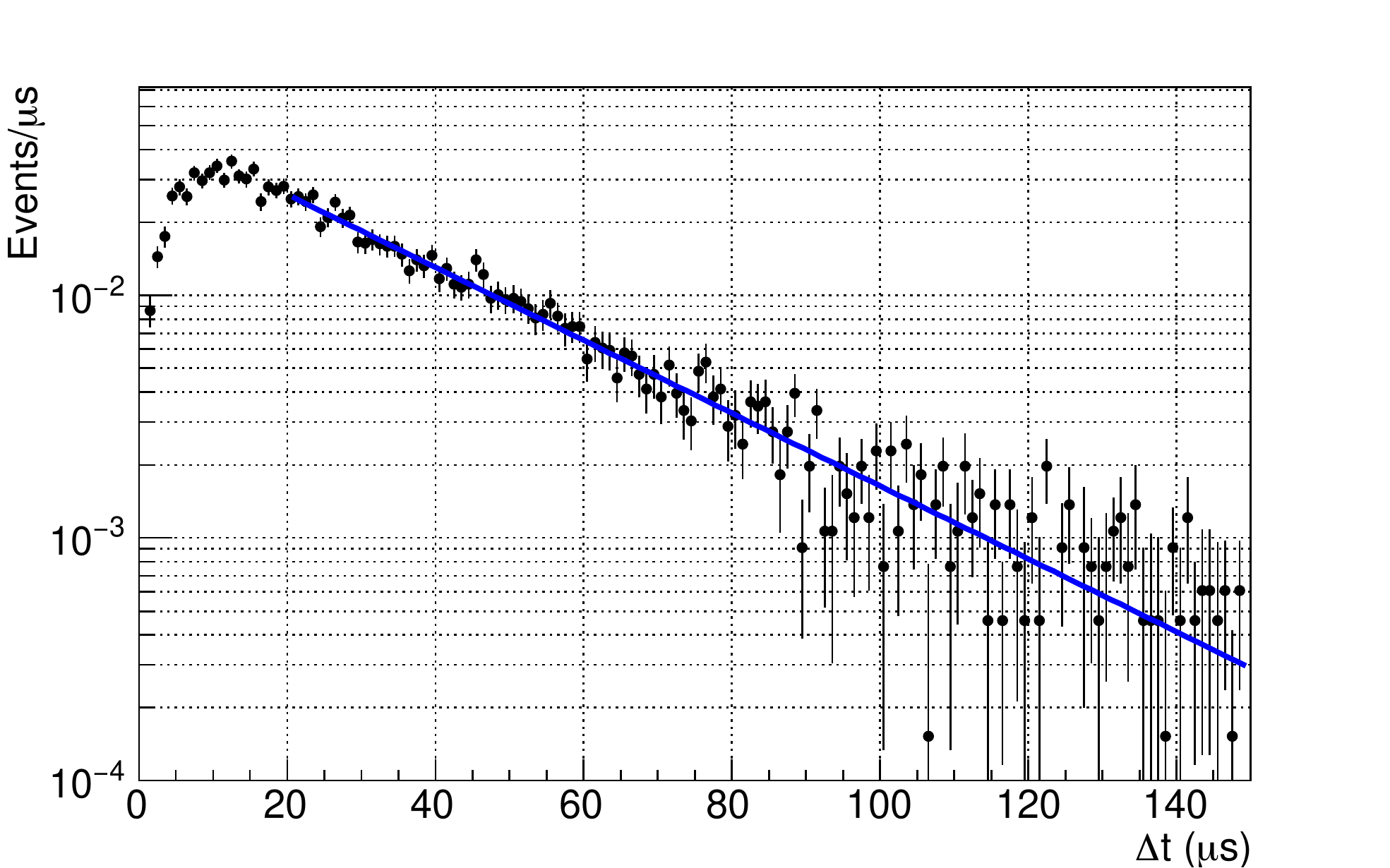}
    \quad
    \includegraphics[width=.45\textwidth]{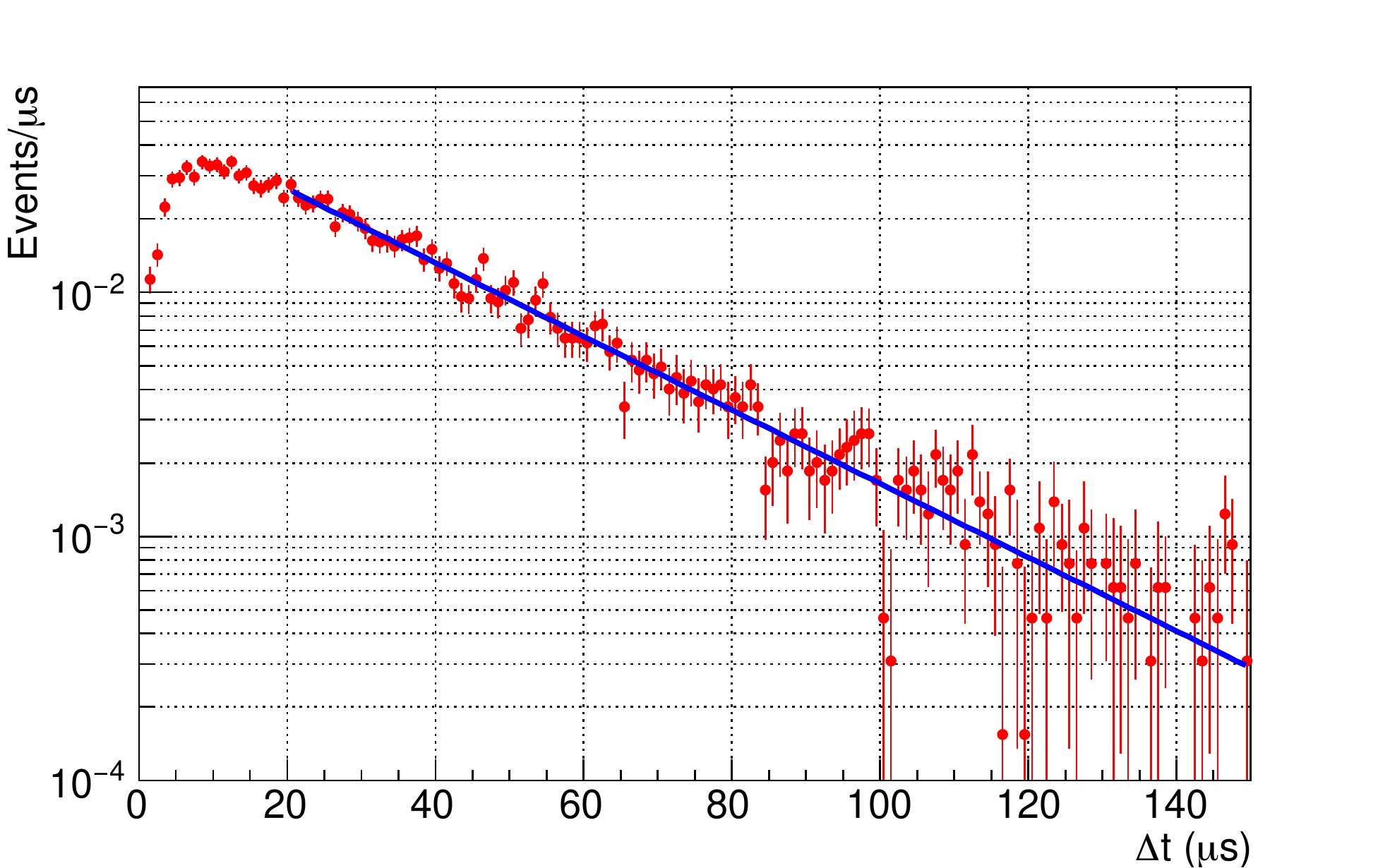}
    \caption{Plots of the fit results of the $\Delta t$ distribution. The blue lines are the best fit curves for each. The fit results show the consistent value within the fit error. \label{fig:CapTfit}}
\end{figure}

The neutron capture time by Gd is evaluated as a time constant by fitting the $\Delta t$ distribution with an exponential function in the range between 20 and 150 \micro (\Figure \ref{fig:CapTfit}).
The result of the neutron capture time measurement exhibited in \Table \ref{tab:CapTResult} shows that the measured capture times are consistent with each other. The measurements were performed in the identical setup; therefore, the systematics for the comparison between the measurements are negligibly small. Thus, one can conclude that 617 days exposure to stainless steel gave no significant degradation on the capture time, which indicates no change of Gd concentration in the Gd-LS.

\begin{table}[hbt]
  \caption{The result of the neutron capture time measurement of the Gd-LS. \label{tab:CapTResult}}
  \centering
    \begin{tabular}{|ccc|}
        \hline
        Exposure /day & Capture Time $\Delta t$ /\micro & $\chi^2/\mathrm{ndf}$ \\ \hline
        0   & $28.92 \pm 0.58$  & 110.8/128 \\ %\hline
        617 & $28.79 \pm 0.58$  & 100.8/128 \\ \hline
    \end{tabular}
\end{table}

\section{EDTA Titration}
In addition to the capture time measurement, we utilised an Ethylene Diamine Tetra Acetic acid (EDTA) titration method~\cite{Danilov:2007} to measure Gd concentration of the Gd-LS. This method was applied to the Gd-LS in RENO experiment~\cite{Park:2013}.

EDTA is one of the well-known chelating agents and forms a chelate complex with a variety of metal ions including Gd. In case of Gd, EDTA makes a chelating complex in the ratio of 1 to 1.
Xylenol Orange (XO), an indicator reagent for metal titrations, was added in order to know the end point of the EDTA titration. XO gives red-violet color in pH $<$ 6 when a metal ion exists. Otherwise, the color shows yellow. Because a chelating complex with EDTA is inactive to XO, the color changes red-violet into yellow gradually as EDTA is dropped. An equivalent point where the amount of EDTA equals that of Gd is determined as the point where the color (yellow) does not change any more.

A micro-pipette was used for dropping 0.01 mol/L EDTA solution and sampling 1 mL Gd-LS to be used in titration. The extraction volume of liquid can be fixed using a dial in 0.01 mL step. The actual volume of the set value is calibrated as explained below.

\subsection{Calibration}
The dominant systematic uncertainty is the total amount of dropped EDTA volume. The uncertainty from the sampling volume of the micro-pipette piles up every drop of EDTA, sampling the Gd-LS.
Thus, liquid volume sampled by the micro-pipette was calibrated using ultra pure water. The sampled volume in a setting was computed from the weight of the sampled water measured by an electronic balance with 1 mg precision.

\begin{table}[htb]
    \caption{Summary of the sampled volume calibration using ultra pure water. The investigation of 1 mL sampling is for a calibration of the sampled volume of the Gd-LS and Gd standard solution. The others are used for EDTA volume calibration. \label{tab:vcalib}}
  \centering 
    \begin{tabular}{|ccc|}
        \hline
        Set Vol. $V_s$/mL & Measured Vol. Mean $V_{\mu}$/mL & Std. Dev. $\sigma_{V}$/mL \\ \hline
        1.00             & 0.985               & 0.005 \\ %\hline
        0.10             & 0.100               & 0.002 \\ %\hline
        0.05             & 0.051               & 0.003 \\ %\hline
        0.01             & 0.015               & 0.002 \\ \hline
    \end{tabular}
\end{table}

Table \ref{tab:vcalib} shows the result of the sampled volume calibration for the micro-pipette.
The mean value of the sampled volume is used as the corrected volume at the setting and the standard deviation is considered as the uncertainty of a drop, respectively.
As a cross-check of the titration with this correction and uncertainty calculation, a certified \ce{Gd(NO_3)_3} standard solution (Wako Pure Chemical Industries) containing $1.000 \pm 0.001 ~ \mathrm{mg/mL}$ of Gd was used.
The result of the titration to the standard solution is exhibited in \Table \ref{tab:StdResult}. We iterated the same titration process 5 times, and each result is in good agreement with the expected Gd $0.985 \pm 0.005$ mg. 

\begin{table}[htb]
    \caption{Results of the titration to the Gd standard solution. The Gd standard solution was sampled 1 mL using the micro-pipette as well; therefore, the expected Gd amount is $0.985 \pm 0.005$ mg. The results of each titration are consistent with the expected value within the estimated uncertainty. \label{tab:StdResult}}
  \centering
    \begin{tabular}{|cccc|}
        \hline
        Measurement & Total Set Vol./mL & Total EDTA Vol./mL & Gd /mg \\ \hline
        1   & 0.59             & $0.635 \pm 0.008$      & $0.999 \pm 0.013$ \\ %\hline
        2   & 0.59             & $0.635 \pm 0.008$      & $0.999 \pm 0.013$ \\ %\hline
        3   & 0.58             & $0.620 \pm 0.008$      & $0.976 \pm 0.012$ \\ %\hline
        4   & 0.58             & $0.620 \pm 0.008$      & $0.976 \pm 0.012$ \\ %\hline
        5   & 0.58             & $0.620 \pm 0.008$      & $0.976 \pm 0.012$ \\ \hline
    \end{tabular}
\end{table}

\subsection{Result}
We measured the Gd concentration of the Gd-LS on three dates. The time interval from the first day are 208 and 466 days respectively.
The result of the titrations is shown on \Table \ref{tab:GdLSResult}. The Gd concentrations of the Gd-LS at each day are consistent with each other within the estimated uncertainty. Thus, there is no significant aging effect caused by the stainless steel on Gd concentration in 466 days. This is consistent with the result of the capture time measurement, which shows no clear change after the long term exposure.

\begin{table}[hbt]
  \caption{The result of the EDTA titration to the Gd-LS. \label{tab:GdLSResult}}
  \centering
    \begin{tabular}{|ccc|}
        \hline
        Measurement Interval /day & Total EDTA Vol./mL & Gd Concentration /w\% \\ \hline
        0   & $0.491 \pm 0.007$ & $0.090 \pm 0.002$ \\ %\hline
        208 & $0.505 \pm 0.007$ & $0.093 \pm 0.002$ \\ %\hline
        466 & $0.498 \pm 0.007$ & $0.091 \pm 0.002$ \\ \hline
    \end{tabular}
\end{table}

\begin{figure}[htbp]
    \centering 
    \includegraphics[width=.6\textwidth]{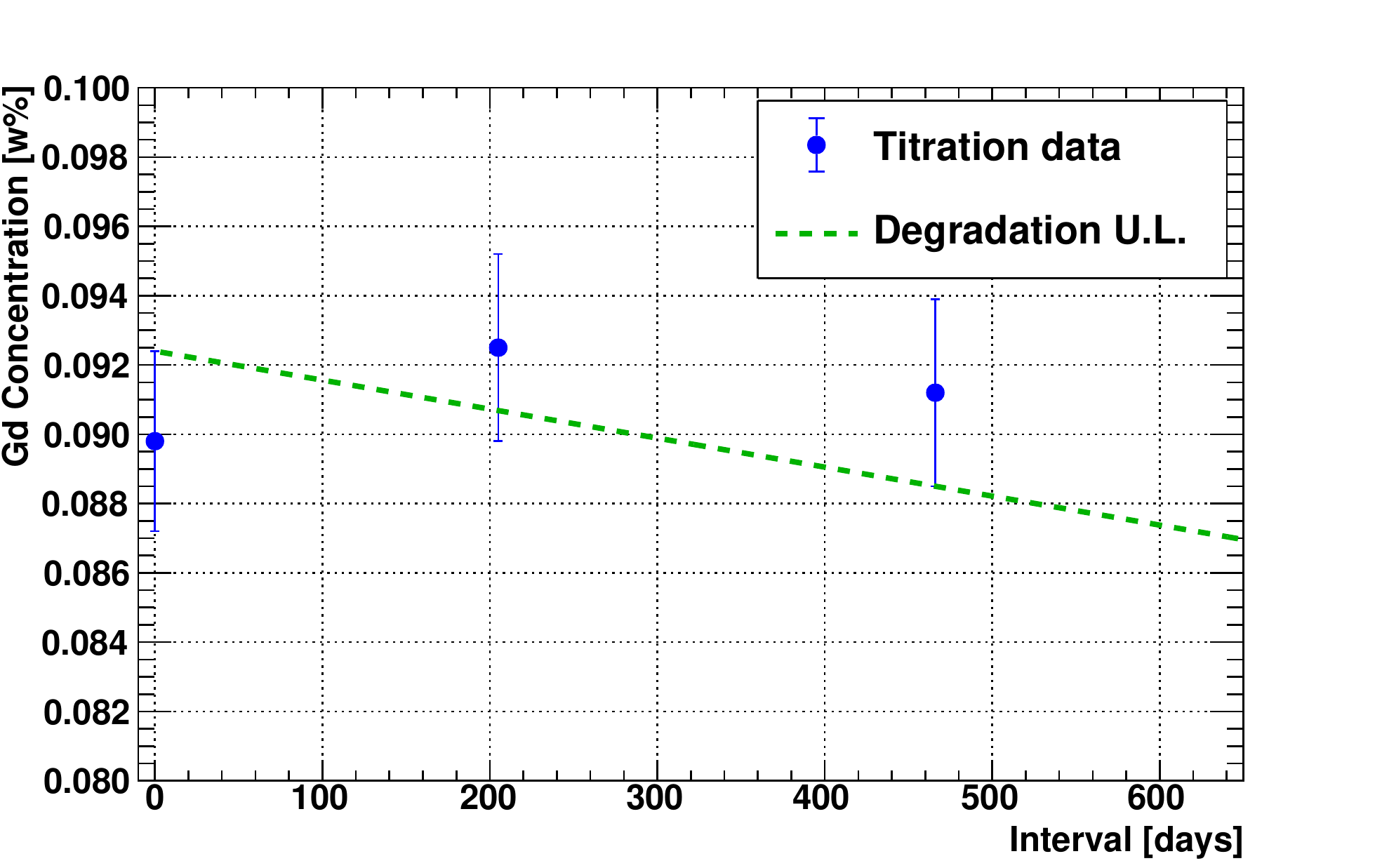}
    \caption{A plot of the titration result. The green line indicates the possible worst degradation within the uncertainties in an assumption. \label{fig:UPL}}
\end{figure}

The upper limit degradation is shown in \Figure \ref{fig:UPL} as the dashed green line with an assumption that the the degradation is linearly proportional to exposure time. This expresses $<$ 0.005 w\% Gd concentration loss at 600 days exposure in the stainless steel bottle. The exposure days, 600 days, is equivalent to the total storage duration in the ISO tank if \JSNS experiment continues for 10 years.
Given the area to volume ratio, therefore, the upper limit of the Gd concentration degradation in the ISO tank is estimated as 0.5 \% with respect to Gd concentration 0.1 w\% for 10 years, which is negligibly smaller than other systematic uncertainties of \JSNS experiment.

\section{Conclusion}
Degradation of Gd concentration of the LAB-based Gd-LS due to exposure to stainless steel vessel is investigated by using two completely different methods. 
The neutron capture time measurement shows that 617 days exposure gave no clear degradation on the neutron capture time by Gd within 2 \% uncertainty.
The results of EDTA titration shows no loss of Gd within the investigated systematics, which agreed with the conclusion of the capture time measurement. The estimated upper limit of the degradation in the ISO tank is 0.5 \% in total experimental duration.
These results reveal that passivated stainless steel surface gives negligibly small impact on Gd concentration in LAB-based Gd-LS in our case, and feasibility of ISO tank storage.

\section*{Acknowledgment}
This work was supported by the JSPS grants-in-aid (Grant Number 16H03967), Japan.
We specially appreciate to Dr.~J.~S.~Park (High Energy Accelerator Research Organization, KEK) for his help and instruction in performing the EDTA titration.

\end{document}